\documentclass[%
 reprint,
 superscriptaddress,
 amsmath,amssymb,
 aps,
]{revtex4-2}

\usepackage{graphicx}
\usepackage{dcolumn}
\usepackage{bm}
\usepackage{xspace}
\usepackage{cleveref}
\usepackage{color}
\usepackage{xcolor}
\usepackage{mathrsfs}

\newcommand{\gev}{\ensuremath{\,\text{GeV}}\xspace}

\mathchardef\-="2D

\begin{document}

\title{
Implications of Nano-Hertz Gravitational Waves on Electroweak Phase Transition in the Singlet Dark Matter Model 
}

\author{Yang Xiao}
\affiliation{CAS Key Laboratory of Theoretical Physics, 
Institute of Theoretical Physics, Chinese Academy of Sciences, Beijing 100190, P. R. China}
\affiliation{School of Physical Sciences, University of Chinese Academy of Sciences, Beijing 100049, P. R. China}
\author{Jin Min Yang}
\affiliation{CAS Key Laboratory of Theoretical Physics, 
Institute of Theoretical Physics, Chinese Academy of Sciences, Beijing 100190, P. R. China}
\affiliation{Department of Physics, Henan Normal University, Xinxiang 453007,  P. R. China}
\author{Yang Zhang}
\email[]{zhangyangphy@zzu.edu.cn}
\affiliation{School of Physics, Zhengzhou University, Zhengzhou 450000, P. R. China}
\affiliation{CAS Key Laboratory of Theoretical Physics, 
Institute of Theoretical Physics, Chinese Academy of Sciences, Beijing 100190, P. R. China}

\date{\today}

\begin{abstract}

Inspired by the recent evidences of nano-Hertz stochastic gravitational waves observed by the pulsar timing array collaborations, we explore their implied supercooled electroweak phase transition in the singlet extension of the Standard Model. Our findings reveal that by adjusting the model parameter at per milli level, the corresponding percolation temperature can be continuously lowered to 1 GeV. With such a low percolation temperature, the singlet dark matter may freeze out before the electroweak phase transition, and, consequently, the entropy generated during the transition can significantly affect the dark matter relic density. It alleviates the tension between the requirement of a strong electroweak phase transition and the constraints imposed by dark matter direct detection, and can be tested in future experiments.\\

\end{abstract}

\maketitle

\section{Introduction}

Recently, the North American Nanohertz Observatory for Gravitational Waves~(NANOGrav), European Pulsar Timing Array~(EPTA), Parkes Pulsar Timing Array~(PPTA), and Chinese Pulsar Timing Array~(CPTA) collaborations reported positive evidences for the presence of stochastic gravitational wave (GW) background in the $\mathcal{O}(1\sim10)$ nHz frequency band~\cite{Xu:2023wog,Reardon:2023gzh,Antoniadis:2023ott,NANOGrav:2023gor}. 
This background can be produced through a variety of cosmological processes~\cite{Bai:2023cqj,Kitajima:2023cek,Yang:2023aak,Megias:2023kiy,Guo:2023hyp,Han:2023olf,Deng:2023seh,Bian:2022qbh,Xue:2021gyq, kitajima2023nanohertz, lazarides2023superheavy,yang2023implication, blasi2023axionic, addazi2023pulsar, broadhurst2023binary,Wang:2022akn,Wang:2022lxn,madge2023primordial,King:2023cgv,Chen:2019xse,Niu:2023bsr,antoniadis2023second, lu2023nanohertz, Huang:2023chx,Jiang:2023gfe,Zhu:2023lbf,Wang2023ost,Cai:2023dls}.
In a model-independent Bayesian analysis of the NANOGrav data, a cosmological phase transition with a percolation temperature around 1 GeV is favored~\cite{NANOGrav:2023hvm}.
However, in popular new physics models beyond the Standard Model, the electroweak phase transition~(EWPT) occurs at around 100 GeV and concludes rapidly, resulting in a milli-Hertz stochastic GW background~\cite{Pietroni:1992in,Cline:1996mga,Ham:2004nv,Funakubo:2005pu,Barger:2008jx,Chung:2010cd,Espinosa:2011ax,Chowdhury:2011ga,Gil:2012ya,Carena:2012np,No:2013wsa,Dorsch:2013wja,Curtin:2014jma,Huang:2014ifa,Profumo:2014opa,Kozaczuk:2014kva,Jiang:2015cwa,Curtin:2016urg, Vaskonen:2016yiu,Dorsch:2016nrg,Huang:2016cjm,Chala:2016ykx,Basler:2016obg,Beniwal:2017eik,Bernon:2017jgv,Kurup:2017dzf,Andersen:2017ika,Chiang:2017nmu,Dorsch:2017nza,Beniwal:2018hyi,Alves:2018jsw,Bruggisser:2018mrt,Athron:2019teq,Kainulainen:2019kyp,Bian:2019kmg,Li:2019tfd,Chiang:2019oms,Xie:2020bkl,Azatov:2022tii,Bell:2020gug,Han:2020ekm,Ghosh:2022fzp,Cao:2022ocg,Zhao:2022cnn,Chatterjee:2022pxf, ashoorioon2009strong}. 
Therefore, it is difficult to explain the observed nano-Hertz GW signals using EWPT. 
Fortunately, the phase transition can be postponed in the case of supercooling. Generally, the percolation temperature, at which the majority of true vacuum bubbles collide, is typically lower than the nucleation temperature by no more than 10 GeV ~\cite{Baratella:2018pxi,Lewicki:2020jiv,Wang:2020jrd,Athron:2022mmm}. The study in \cite{Kobakhidze:2017mru} found that the percolation temperature can descend to a few MeV, with a nucleation temperature of approximately 50 GeV, in a toy model that is based on a non-linear realization of the electroweak gauge group. A similar type of extremely supercooled first-order phase transition~(FOPT) is investigated in \cite{Cai:2017tmh} within the framework of the SM extended with a dimension-six operator.

In this study, we investigate the phenomenon of extreme supercooling within a more realistic model, i.e., the singlet extension of the SM with $\mathbb{Z}_2$ symmetry (SSM), which contains a dark matter~(DM) candidate. This model is highly restricted by the DM direct detection limits~\cite{Cline:2013gha,GAMBIT:2017gge}.
Even when taking into account the dilution effect caused by the supercooled phase transition, these constraints cannot be alleviated as the freeze-out temperature is lower than the nucleation temperature~\cite{Xiao:2022oaq,Roy:2022gop}.
Nonetheless, inspired by the observed evidence of nano-Hertz stochastic GWs, it is possible that the EWPT ends at a temperature of a few GeV. In this scenario, the freeze-out of DM may occur before the completion of the phase transition, and thus the DM density can be diluted by entropy release during the strong first-order phase transition~\cite{Hambye:2018qjv,Baldes:2021aph,Xiao:2022oaq}.  

By and large, it is possible to generate the reported nano-Hertz stochastic GWs through an extremely supercooled EWPT in new physics models. Accordingly, the relevant phenomenology of DM needs to be revisited, as the DM decouples from other particles during the EWPT, which may have significant implication for the abundance of DM.

The work is organized as follows. In Section II, we provide an introduction to our model and discuss the physics associated with the phase transition. Section III shows the range of nucleation temperature and percolation temperature in the SSM, and demonstrates the corresponding spectrum of GWs. In Section IV, we analyze the implications of a low percolation temperature on the calculations of dark matter. Finally, we summarize our findings and draw our conclusion in Section V. 

\section{Singlet extension of the SM}

The SSM is one of the simplest and most predictive realisations of the weakly interacting DM scenario. In this model, the addition of an extra scalar field allows for the generation of a potential barrier between the high-temperature symmetric minimum and the electroweak symmetry breaking (EWSB) minimum as the universe cools down. This results in a strong first-order EWPT, which has the potential to generate the observed baryon asymmetry of the universe and produce detectable stochastic GWs  (see \cite{Caprini:2019egz,Athron:2023xlk} for recent reviews). 

After some parameterization, the tree-level effective potential of the SSM can be expressed as
\small 
\begin{equation}\label{eq:xsm_tree}
     V_{0}(\phi_{\rm h},\phi_{\rm s}) = -\frac{\mu_{\rm h}^{2}}{2}\phi_{\rm h}^2 + \frac{\lambda_{\rm h}}{4}\phi_{\rm h}^4 - \frac{\mu_{\rm s}^2}{2}\phi_{\rm s}^2 + \frac{\lambda_{\rm s}}{4}\phi_{\rm s}^4 + \frac{\lambda_{\rm hs}}{4}\phi_{\rm h}^2\phi_{\rm s}^2,
\end{equation}
\normalsize 
where $\phi_{\rm h}$ and $\phi_{\rm s}$ represent the background field configurations for the SM Higgs and the additional scalar field, respectively. The model parameters satisfy the tadpole conditions,
\begin{equation}\label{eq:xsm_tadpole}
\begin{aligned}
      \left. \frac{\partial V_0}{\partial \phi_{\rm h}} \right|_{{v}}  = 0&, ~~~~ \left. \frac{\partial V_0}{\partial \phi_{\rm s}} \right|_{{v}}   = 0 , \\
      ~~ \left. \frac{\partial^2 V_0}{\partial \phi_{\rm h}^2} \right|_{{v}}   = m_{\rm h}^2 &, ~~~~ \left. \frac{\partial^2 V_0}{\partial \phi_{\rm s}^2} \right|_{{v}}   = m_{\rm s}^2,
\end{aligned}
\end{equation}
at the electroweak vacuum $v \equiv (v_{\rm EW}, 0)$. Here we set $m_{\rm h}=125~\gev$ and $v_{\rm EW} = 246~\gev$. As a result, there remains three free parameters, namely $m_{\rm s}$, $\lambda_{\rm s}$, and $\lambda_{\rm hs}$. For simplicity, we incorporate the one-loop correction using the on-shell-like renormalization scheme in Landau gauge, which maintains the above tadpole conditions. The total effective potential is given by
\begin{eqnarray}\label{eq:xsm_potential}
     V(\phi;T) =  V_0(\phi) + V_1(\phi;T) + V_{\rm 1T}(\phi;T),
\end{eqnarray}
where 
\begin{eqnarray}
V_1(\phi;T) &=&  \frac{1}{64\pi^2} \sum_{k} (-1)^{2s_k} n_{k} \left\{  \rule{0pt}{10pt} m_{k}^4(\phi;T) \right. \nonumber \\
&&   \times \left[\log\frac{m_{k}^2(\phi;T)}{m^2_{k}(v;0)}  
       -\frac{3}{2}\right]  
 \nonumber \\
&&  \left.        
       + 2 {m_{k}^2(\phi;T)}{m^2_{k}(v;0)} - \frac{1}{2} {m^4_{k}(v;0)}\right\} ,\\
      V_{\rm 1T}(\phi;T)  &=&  \frac{T^4}{2\pi^2} \sum_{k} n_k J_{\rm B,F}\left[ \frac{m_{k}(\phi;T)^2}{T^2}\right].
\end{eqnarray}
Here $\phi$ represents $(\phi_{\rm h},\phi_{\rm s})$, $J_{\rm B,F}$ are the thermal functions for boson and fermion, $m_{k}(\phi;T)$ is the field-dependent thermal mass including Debye corrections, $k$ ranges over the entire field-dependent mass spectrum except Goldstone bosons to remedy the infrared divergences, $n_k$ and  $s_k$ represent the corresponding degrees of freedom and spin, and the Parwani method~\cite{parwani1992resummation} is adopted for daisy resummation. 
The resummed effective theory enables the computation of advanced state-of-the-art calculations~\cite{Croon_2021, Niemi_2021, Schicho_2021}.
The Mathematica package \texttt{DRalgo} can be used to perform these computations~\cite{Ekstedt_2023}.

In general, altering the settings in the effective potential would have a tolerable impact on the properties of EWPT, except when the transition temperature is sensitive to the model parameters~\cite{athron2022arbitrary}. 
Unfortunately, the situation of extreme supercooling is highly sensitive to the model parameters. Consequently, even slight changes to the above settings can lead to different percolation temperatures. However, we find that the change in percolation temperature with varying $\lambda_{\rm hs}$ is continuous, unlike the behavior of the nucleation temperature. This implies that we can always tune the model parameters to achieve a similar result shown below.

In the evolution of the effective potential at finite temperature, there are three commonly used temperatures to characterize the process of phase transition: the critical temperature $T_{\rm c}$, the nucleation temperature $T_{\rm n}$, and the percolation temperature $T_{\rm p}$.  

The critical temperature $T_{\rm c}$ is defined as the temperature at which the two minimums become degenerate,
\begin{equation}
     V(v_{\rm h}^{\rm high},v_{\rm s}^{\rm high};T_{\rm c}) = V(v_{\rm h}^{\rm low},v_{\rm s}^{\rm low};T_{\rm c}),
\end{equation}
where $V(\phi_{\rm h},\phi_{\rm s},T)$ is the full one-loop finite temperature effective potential, and the minimums of $(v_{\rm h}^{\rm high},v_{\rm s}^{\rm high})$ and $(v_{\rm h}^{\rm low},v_{\rm s}^{\rm low})$ correspond to the high-temperature symmetric minimum and the low-temperature EWSB minimum, respectively. In the SSM, we have $v_{\rm h}^{\rm high}=0$ and $v_{\rm s}^{\rm low}=0$, while $v_{\rm s}^{\rm high}$ usually increases continuously from zero and $v_{\rm h}^{\rm low}$ approaches  $v_{\rm EW}$ at zero-temperature. 

When the temperature of the universe falls below $T_{\rm c}$, the low-temperature EWSB minimum starts to have lower free energy than the high-temperature symmetric minimum. Thus some regions of the symmetric plasma tunnel to the true vacuum with a probability per unit volume per unit time ~\cite{linde1983decay, Kobakhidze:2017mru}:
\begin{equation}
    \Gamma \sim A~e^{-S},
\end{equation}
where $S$ is given by
\small 
\begin{align}\label{eq:action}
\begin{split}
 S= \left \{
 \begin{array}{ll}
 2 \pi^2 \int^{+\infty}_{0}r^3{\rm d}r~\left[\frac{1}{2}\left(\frac{\partial \phi}{\partial r}\right)^{2} + V_{\rm eff}(\phi;T)\right],                    & T \approx 0\\
 \frac{4 \pi}{T} \int^{+\infty}_{0}r^2{\rm d}r~\left[\frac{1}{2}\left(\frac{\partial \phi}{\partial r}\right)^{2} + V_{\rm eff}(\phi;T)\right],     & T \gg 0.\\
 \end{array}
 \right.
 \end{split}
 \end{align}
\normalsize 
Here the bubble configuration $\phi(r)$ in the integral is fixed from the corresponding equation of motion
\begin{equation}
    \frac{{\rm d}^2 \phi}{{\rm d} r^2} + \frac{d-1}{r}\frac{{\rm d} \phi}{{\rm d}r} = \frac{\partial V_{\rm eff}(\phi;T)}{\partial \phi},
\end{equation}
subjected to the boundary conditions $\lim \limits_{r \to \infty} \phi(r) = 0 $ and ${\rm d} \phi/{\rm d} r|_{r=0}=0$~\cite{rubakov2009classical, linde1983decay}.
The pre-factor $A$ is often estimated as the fourth power of the temperature  when temperature is high and the fourth power of the energy scale when temperature is zero. In this paper, we set $A$ to $T_{\rm c}^4 \sim \mathcal{O}(100~\gev)^4$ as in \cite{guth1981cosmological,hindmarsh2019gravitational},
for all the temperature, as the EWPT energy scale is the same order as $T_{\rm c}$. 

The nucleation rate of bubbles increases significantly as the universe continues to cool. The phase transition begins when the probability of nucleating a supercritical bubble within one Hubble volume becomes approximately one, which gives the definition of $T_{\rm n}$: 
\begin{equation}
    \int^{+\infty}_{T_{\rm n}} \frac{{\rm d} T}{T}\frac{\Gamma(T)}{H(T)^4} = \mathcal{O}(1),
\end{equation}
where $H(T) = \sqrt{8\pi G \rho/3}$ is the Hubble constant, $G$ is the gravitational constant, and $\rho$ is the energy density of the universe~\cite{quiros1998finite}. 
From this definition, we can get an approximate formula for $T_{\rm n}$,
\begin{equation} \label{eq: S/T}
    S \approx 4~{\rm ln}\frac{M_{\rm Pl}}{T} \approx 130 \sim 140 ,
\end{equation}
where the Planck scale is $M_{\rm Pl}=1.22\times 10^{19}$ GeV. 

With the temperature further decreasing, the nucleated bubbles of true vacuum keep growth and occupy nearly 30\% of the space when the percolation temperature $T_{\rm p}$ is reached. This percentage is determined by the formation of a cluster of connected bubbles with size of the order of the medium, i.e., bubbles are colliding~\cite{Athron:2022mmm}. Therefore, $T_{\rm p}$ is crucial for the stochastic GW background produced from bubble collision. 

The calculation of $T_{\rm p}$ involves approximating the fraction of false vacuum~\cite{guth1981cosmological},
\begin{equation} \label{eq: h(t)}
    h(t) = {\rm{exp}}[-\int^t _{t_{\rm initial}}\Gamma(t')V(t',t){\rm d}t'], 
\end{equation}
where $v_{\rm w}$ is the bubble velocity and 
\begin{equation} 
V(t',t) = g \left[\int ^t _{t'} v_{\rm w}(\tau) {\rm d}\tau\right]^3. 
\end{equation} 
For a spherical bubble, the shape constant $g$ is equal to $4\pi/3$. 
In general, the fraction of false vacuum undergoes a significant change around the percolation temperature $T_{\rm p}$. Therefore, the accuracy of the computational results relies on the stability of the action calculation. Nonetheless, in the case of the SSM, the stability of the action calculation is not satisfactory in \texttt{CosmoTransitions}~\cite{wainwright2012cosmotransitions}. Thus, we repeat the calculation of $T_{\rm p}$ for each sample, find the interval that includes the percolation temperature, and ensure that the length of the interval is small enough to safely consider the average value as the percolation temperature.

In a fast phase transition, these three temperatures are closely aligned with one another. However, in a supercooled transition, they become noticeably separate, resulting in an enlargement of the energy gap as the transition progresses.

\section{Results and discussions}

\begin{figure}[t]
\centering 
\includegraphics[width=0.49\textwidth]{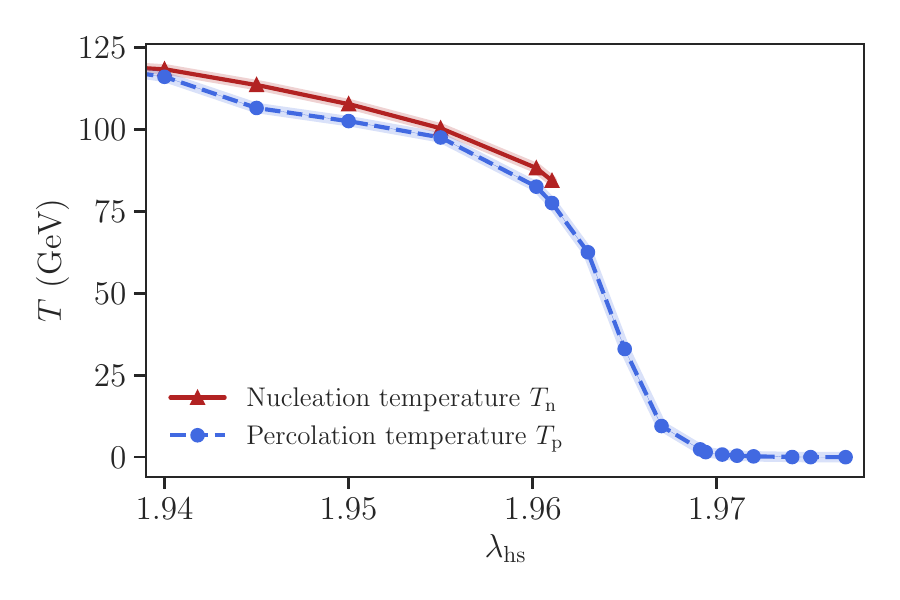}
\caption{(Color online) The nucleation temperature  and the percolation temperature versus the mixing coupling $\lambda_{\rm hs}$ in the SSM. Other parameters are fixed using the tapole conditions with fixed $m_{\rm s}=234~\gev$ and $\lambda_{\rm s}=0.2$. }
\label{fig:TnTp}
\end{figure}

In Fig.~\ref{fig:TnTp} we present the nucleation temperature and the percolation temperature for a set of benchmark points in the SSM ($m_{\rm s}=234~\text{GeV}$, $\lambda_{\rm s}=0.2$, $\lambda_{\rm hs}\simeq 1.96$). These particular points are selected near the line where the two phases (with zero or nonzero $v_{\rm EW}$) become degenerate at zero temperature, i.e. $T_{\rm c} = 0$, satisfying~\cite{Balazs:2023kuk}
\begin{equation}
    \frac{1}{2} \lambda_{\rm hs} v_{\rm EW}^2 - \mu_{\rm h}^2\sqrt{\frac{\lambda_{\rm s}}{\lambda_{\rm h}}} = m_{\rm s}^2.
\end{equation}
On this line, there will be no EWPT at all. \
By tuning the model parameters, such as by decreasing $\lambda_{\rm hs}$, we can achieve an EWPT at a significantly low temperature. It is worth noting that the transition temperature is highly sensitive to the model parameters when it is small~\cite{athron2022arbitrary}.  We start our scan from this line, using \texttt{EasyScan\_HEP}~\cite{Shang:2023gfy}. 
The results are illustrated in Fig.~\ref{fig:TnTp}. In addition to the data points obtained from the grid scan, we have included the specific endpoint of the $T_{\rm n}$ curve, as well as several manually selected data points indicating low values of $T_{\rm p}$, which will serve as benchmark points for subsequent analysis.

\begin{figure*}[ht]
\centering 
\includegraphics[width=0.95\textwidth]{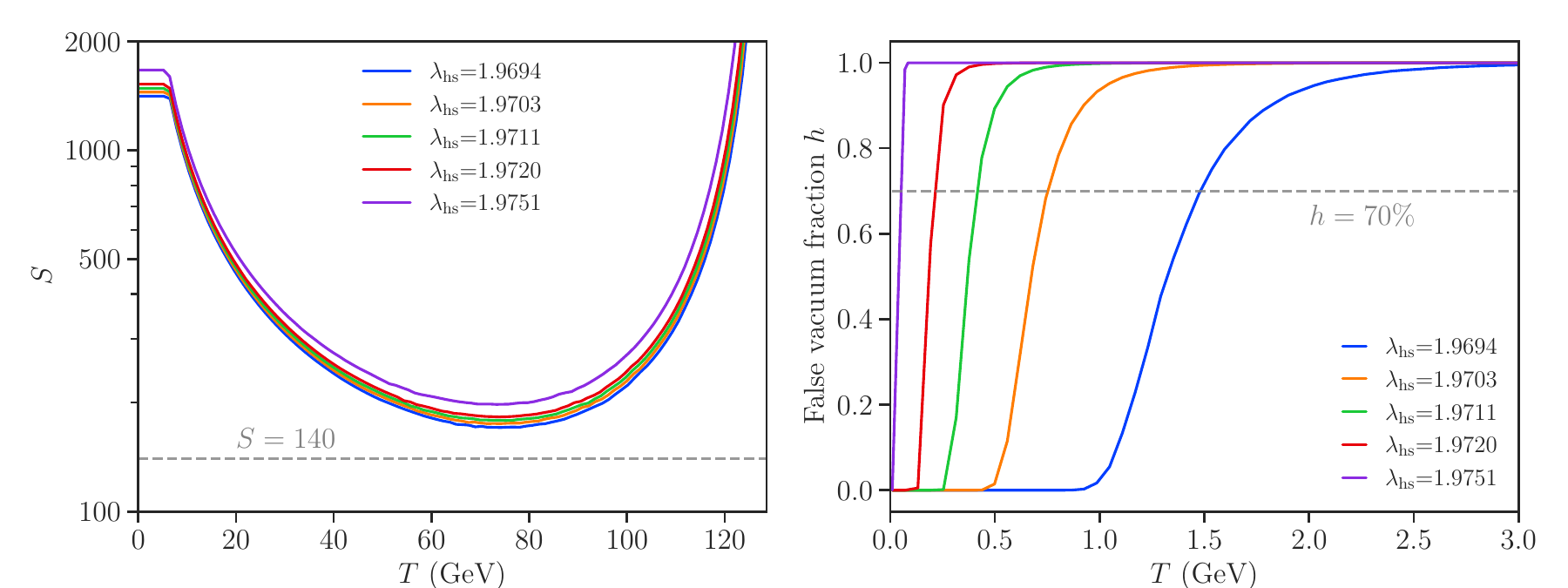}
\caption{(Color online) The $S$ value (left panel) and the false vacuum ratio (right panel) versus the temperature for benchmark points $\lambda_{\rm hs} \in [1.9694, 1.9751]$, where other parameters are fixed as $m_{\rm s}=234~\gev$ and $\lambda_{\rm s}=0.2$. In the left panel, the lines from top to bottom correspond to increasing lambda values. In the right panel, the lines from left to right correspond to decreasing lambda values.}
\label{fig:bks}
\end{figure*}

We can see that both $T_{\rm n}$ and $T_{\rm p}$ decrease as $\lambda_{\rm hs}$ increases, because a smaller $\lambda_{\rm hs}$ leads to a smaller energy gap between the two minima.  The curve of $T_{\rm n}$ ends at 84~\gev with $\lambda_{\rm hs} = 1.961$. The reason of such an end can be seen from the left panel of Fig.~\ref{fig:bks}, or Figure 4 in~\cite{Xiao:2022oaq} and Figure 2 in~\cite{Cai:2017tmh}.  For a given point, as the temperature decreases, the action $S$ initially decreases, but it may start to increase before reaching approximately 140 due to the temperature appearing in the denominator of Eq.~\ref{eq: S/T}. The lower bound for $T_{\rm n}$ in the SSM is approximately 44~GeV~\cite{Xiao:2022oaq,Azatov:2022tii}. In \cite{Azatov:2022tii}, a much lower $T_{\rm n}$ can be achieved in the super fine-tuned region where the high-temperature symmetric mumimum has non-zero $v_{\rm h}^{\rm high}$. Here, we fix the values of $\lambda_{\rm s}$ and $m_{\rm s}$ such that the lowest $T_{\rm n}$ is approximately 84 GeV, to demonstrate that it is still possible to find $T_{\rm p}$ around 1 GeV even when the lower bound on $T_{\rm n}$ is high.

Before the nucleation temperature disappears, the difference between $T_{\rm n}$ and $T_{\rm p}$ is relatively small, around 5~\gev. Then, $T_{\rm p}$ decreases dramatically and continuously from 80~\gev to zero. This means that, without considering any other constraints, we can obtain any desired value of $T_{\rm p}$ by finely tuning $\lambda_{\rm hs}$ at a level lower than 1\textperthousand. Of course, $T_{\rm p}$ should be at least large than 1 MeV to satisfy nucleosynthesis constraints. 

The absence of a nucleation temperature does not imply the absence of bubbles, but indicates that the statistical average number of bubbles within a Hubble volume is less than one. It is still possible for a few bubbles to be generated and expand until they fill the entire universe. As a consequence, the collisions of these bubbles can generate a stochastic GW background.  

\begin{figure}[ht]
\centering 
\includegraphics[width=0.48\textwidth]{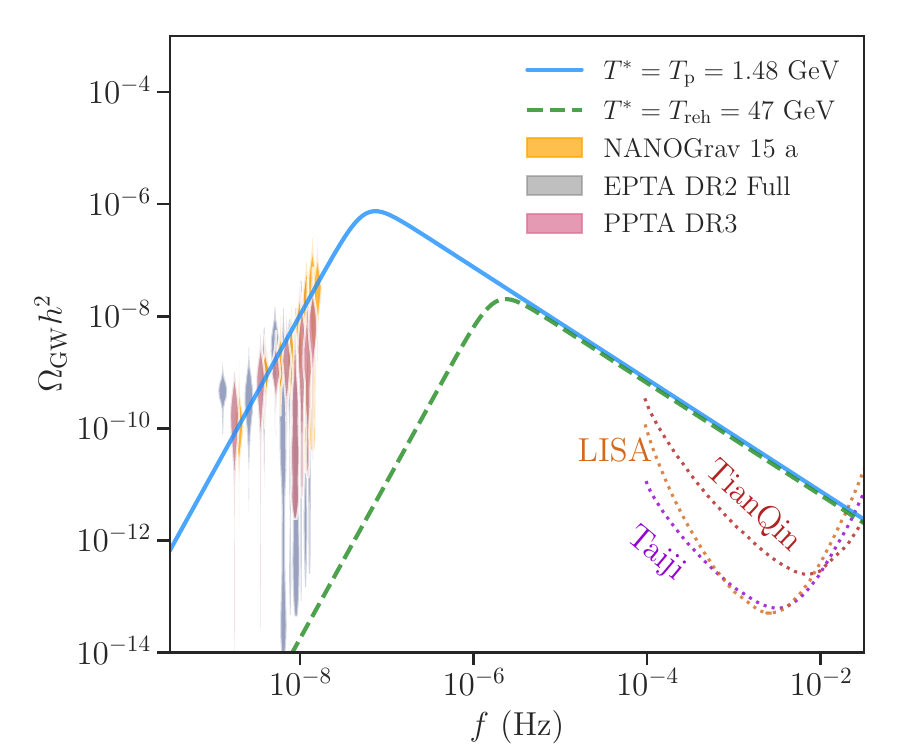}
\caption{(Color online) The spectrum of stochastic GW background generated by bubble collisions in the SSM, for a reference temperature of 1.48~\gev (blue solid curve) and 47~\gev(green dashed curve), with the observations from pulsar timing
arrays (violin plots) and the future detection capabilities (dotted curves).
}
\label{fig:GW}
\end{figure}

Fig.\ref{fig:bks} illustrates the action and the false vacuum rate versus temperature for the benchmark point with $\lambda_{\rm hs}\in[1.9694,~1.9751]$, which corresponds to $T_{\rm p} \in [1.48, 0.02]~\mathrm{GeV}$. The action consistently remains above 140, indicating that there is no nucleation temperature. 
In this situation, only a few bubbles can be generated. Such a low $T_{\rm p}$ indicates that the dominant phase transition mode is not the thermal transition but the quantum tunneling, which can be observed from the flat area in the left panel of Fig.\ref{fig:bks}. Typically,  the dominant source of spectrum of stochastic GW background is the sound wave in EWPT~\cite{hindmarsh2014gravitational}.
However, in the case of extreme supercooling, most of the released energy is utilized to accelerate the bubble walls, making the bubble collision as the dominant source. Additionally, it is safe to approximate the velocity of the bubble wall as 1 instead of solving the Boltzmann equation, which is the value used in the recent NaNoGrav report~\cite{NANOGrav:2023hvm}.

Fig.~\ref{fig:bks} also shows that the supercooled phase transition lasts for a considerable duration, which makes the expansion effect of the universe non-negligible. We estimate that this effect  will reduce the value of $T_{\rm p}$ for the benchmark point of $\lambda_{\rm hs}=1.9694$ from 1.48 GeV to 0.06 GeV, and the transition can still accomplish, using the approximate method for the simplified model~\cite{Kobakhidze:2017mru}. Fortunately, due to the continuous variation of $T_{\rm p}$ with the parameter, it is possible to identify a new point where $T_{\rm p}$ is close to the previously obtained value. By increasing $\lambda_{\rm hs}$ to $1.9671$, we find a new point where $T_{\rm p}$ is approximately 1.6 GeV, accounting for the expansion effect of the universe.

Using the results of~\cite{huber2008gravitational}, the spectrum  of GW generated through bubble collisions can be described as 
\begin{eqnarray}\label{eq:gw1}
    \Omega_{\rm col}h^2 &=& 1.67\times 10^{-5}(\frac{100}{g*})^{1/3}(\frac{\beta}{H_{*}})\kappa_{\rm p}^2(\frac{\alpha}{\alpha+1})^2
    \nonumber \\
    && \times  \frac{0.11v^3}{0.42+v^2} S(f),
\end{eqnarray}
where
\begin{eqnarray}\label{eq:gw2}
    S(f) &= &\frac{3.8(f/f_0)^{2.8}}{1+2.8(f/f_0)^{3.8}},\\
    f_0 &=& 1.65~\times~10^{-7}(\frac{T_{*}}{\gev})(\frac{g*}{100})^{1/6}(\frac{\beta}{H_{*}}) \nonumber\\
    && \times \frac{0.62}{1.8-0.1v+v^2}~\rm{Hz},\\
    \kappa_{\rm p} &=& \frac{1}{1+0.715\alpha}(0.715\alpha+\frac{4}{27}\sqrt{\frac{3\alpha}{2}}),
\end{eqnarray}
with $g*$ being the degree of freedom of relativistic particles, $\alpha$ being the ratio of the vacuum energy to the radiation energy, $v$ being  the bubble wall velocity and $T_{*}$ being the reference temperature. A few remarks are in order: 
\begin{enumerate}
    \item In this extreme supercooling case, $\alpha$ is sufficiently large, exceeding $10^5$ for the benchmark points, so that the ratios involving $\alpha$ in Eq.~\ref{eq:gw1} and Eq.~\ref{eq:gw2} can be regarded as one. The definition of $\kappa_{\rm p}$ here may not be appropriate in our particular case, but it is expected to equal one due to the dominance of vacuum tunneling in the transition process.
    \item The reference temperature is often set as the nucleation temperature or the percolation temperature. Recently, the study in \cite{Athron:2022mmm} proposed that the nucleation temperature may not be suitable for this extreme case, and the percolation temperature can accurately reflect the phase transition process. Therefore, we use the percolation temperature as the reference temperature to avoid the dilemma of non-existence of the nucleation temperature.
    \item The parameter $\beta$ is often defined as the derivative of the thermal action: $\beta/H_* = T{\rm d}(S_3/T)/{\rm d}T$. However, in the case of supercooling, where the dominant action is the 4D action, $\beta$ becomes zero, as shown in Fig. \ref{fig:bks}. Another way to calculate $\beta$ is based on dimensional analysis~\cite{Kobakhidze:2017mru}, where $\beta \sim vR^{-1} \sim R^{-1}$, with $R$ being the characteristic length scale chosen as the radius of the bubble. Thus, $\beta/H_* \sim 1/(RH_*) \sim V^{1/3}/R \sim \mathcal{O}(1)$, due to the fact that the entire universe is occupied by only a few bubbles. This estimation of $\beta$ is consistent with the result of \cite{Kobakhidze:2017mru}. A low bound of $\beta/H_*>3$ is introduced in 
    \cite{Freese:2022qrl} 
    to prevent phase transitions from being incomplete or leading to eternal inflation. It assumes that $\beta/H_*$ is of the same order as $S_3/T$ at nucleation, while our scenario involves $S\simeq S_4$ during the phase transition.
\end{enumerate}

The spectrum of stochastic GW background generated by these collisions for $\lambda=1.9694$ is shown by the blue curve in Fig.~\ref{fig:GW}. The grey band represents the observations from pulsar timing
arrays~\cite{Reardon:2023gzh,Antoniadis:2023ott,NANOGrav:2023gor}~(summarized in \cite{Guo:2023hyp}), while the dotted curves indicate the future detection capabilities from the Laser Interferometer Space Antenna~(LISA, brown)~\cite{amaro2017laser}, Taiji~ (purple)~\cite{ruan2020taiji}, and TianQin~(red)~\cite{luo2016tianqin}, which are taken from \cite{Bian:2021ini,Zhao:2020iew}. We observe that the spectrum associated with $T_{\rm p} \approx 1.48~\gev$ displays a peak frequency that coincides with the NANOGrav signals. 
The results are consistent with the nano-Hertz background produced in the similar way by the one-dimensional effective potentials ~\cite{Kobakhidze:2017mru, Cai:2017tmh}. 

The energy released during the supercooled phase transition will heat up the surrounding plasma and cause a shift in the peak frequency of the stochastic GW background.
In \cite{Athron:2023mer}, the reheating temperature is estimated to be approximately at the energy scale of the new physics, assuming conservation of energy density during the reheating process (this will be further discussed in the next section).
Meanwhile, \cite{Ertas:2021xeh} pointed out the entropy injection of phase transition could lead to a strong dilution of the GW signal.
Consequently, the peak frequency of the background will undergo a red-shift from the desired nano-Hertz range. 
We illustrate this shift by the green dashed-line in Fig.\ref{fig:GW}, with a simplified estimate of $T_{\rm reh} \approx 47\gev$, which is obtained from Eq.~\ref{eq:wain_reh} and will be discussed later. 
This issue cannot be avoided by simply decreasing $T_{\rm p}$ to a much lower value. We have checked numerically that the corresponding latent heat $\alpha$ decreases proportionally to $1/T^4$ in the case of extreme supercooling. Consequently, even if we reduce $T_{\rm p}$ to the MeV scale, the reheating temperature $T_{\rm reh}$ still remains on the order of dozens of GeV. 

In this study, the main focus is to demonstrate the existence of a remarkably low percolation temperature in a concrete model, which serves as a prerequisite for explaining nano-Hertz GW. Further investigations are necessary to verify and understand this complex scenario.

\section{Implications on dark matter}

We previously found in \cite{Xiao:2022oaq} that the dilution effect caused by an electroweak FOPT is negligible for the current DM density in the SSM. This is because the freeze-out temperature $T_{\rm f}$ is always lower than the nucleation temperature, indicating that the strong FOPT typically occurs before the DM freeze-out. In the SSM, the freeze-out temperature can be approximated as $T_{\rm f} \approx m_{\rm s}/20~\gev$, where $m_{\rm s}$ is required to be smaller than 1 TeV for a strong FOPT. Consequently, we have $T_{\rm n} > 50~\mathrm{GeV} > T_{\rm f}$. 

Nevertheless, in this unique situation of extreme supercooling, the phase transition completes when the temperature of the universe drops below $T_{\rm p}$, which is below the GeV scale. This is significantly lower than the freeze-out temperature calculated using the traditional method. Thus, the dilution effect caused by FOPT can be preserved and has an impact on the current DM relic density. This effect can potentially rescue parameter space that was previously excluded by DM direct detection experiments or by an excessive DM relic density. Note that the calculation of dilution factor described in \cite{wainwright2009impact} may not be applicable in this case, as it assumes not very strong supercooling. It also assumes that the energy density is conserved during the reheating as in ~\cite{Athron:2023mer}, but allows a fraction $\xi$ of the universe to be occupied by the true vacuum. Then, we have
\begin{eqnarray} \label{eq:wain_reh}
    \rho(\phi_{\rm f}, T') &=& \rho(\phi_{\rm f}, T_{\rm reh}) - \xi [\rho(\phi_{\rm f}, T_{\rm reh})-\rho(\phi_{t}, T_{\rm reh})] \nonumber \\
    & =& \rho(\phi_{\rm f}, T_{\rm reh}) - \xi L,
\end{eqnarray}
where $\phi_{\rm f}$ and $\phi_{t}$ represnt false and true vacuums, respectively.
We can determine the value of $\xi$ once we know the reheating temperature, or vice versa. For the benchmark point with $\lambda_{\rm hs} = 1.9694$, by setting $f \approx 0.3$ and $T' = T_{\rm p}$, we can find that the corresponding reheating temperature is approximately $47~\text{GeV}$, which is consistent with another estimation method $T_{\rm reh} \sim (1+\alpha)^{1/4}T_{\rm p} = 40.7$ GeV ~\cite{Ellis:2018mja}. 
On the other hand, we can calculate the true vacuum ratio $f$ for the case when $T' = T_{\rm p}$ and $T_{\rm reh}=T_{\rm c}$, where the maximum reheating temperature corresponds to the critical temperature at which the universe is in the 
phase-coexistence stage. This gives  $f \approx 14 \gg 1$, which is clearly unphysical.
Similar results can also arise even in cases where the phase transition is not extremely supercooled, as demonstrated in ~\cite{Xiao:2022oaq}.
It suggests that the assumption of energy density conservation may not hold in this scenario. It is necessary to consider more dynamic processes, as discussed in \cite{megevand2008supercooling}, to better understand this situation. Additionally, the calculation of freeze-out temperature before EWSB is an unsolved problem. Therefore, a specific study is required to determine the dilution factor, and we leave this for future research.

\begin{figure}[t]
\centering 
\includegraphics[width=0.48\textwidth]{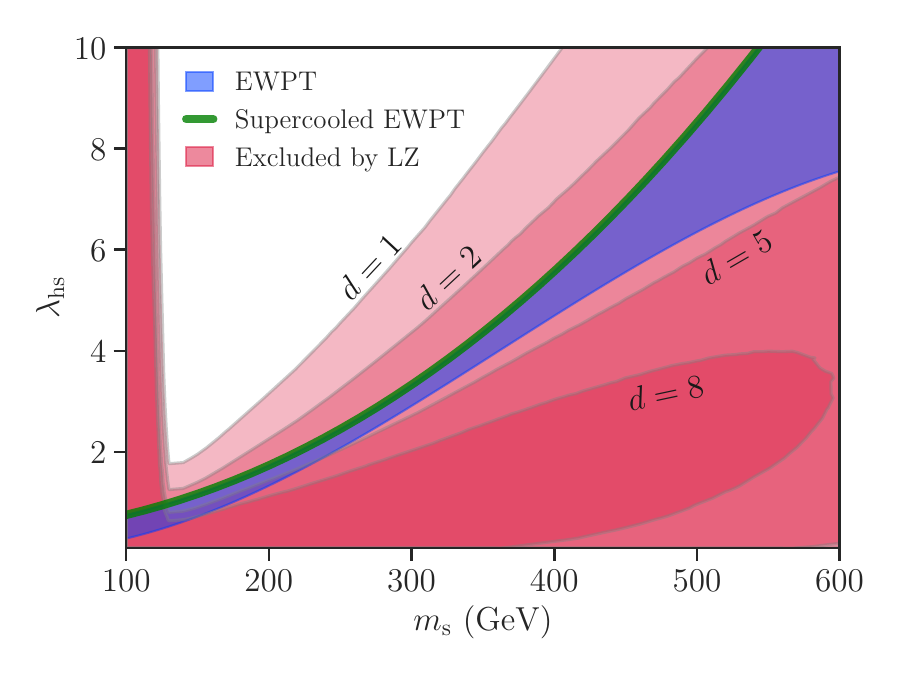}
\caption{(Color online) Limits from DM direct detection on SSM taking into account the dilution effect. The areas excluded by current limits from LZ are delineated with red shading, with colors from light to dark corresponding to dilution factor $d=1, 2, 5, 8$, respectively. The blue band indicates the parameter space where a strong EWPT occurs, while the transition is supercooled near the green line.}
\label{fig:DM}
\end{figure}

Fig.~\ref{fig:DM} displays the limits from DM direct detection on the parameter space of SSM where a supercooled EWPT may occur. We assume that the singlet scalar $s$ only contributes a fraction $f_{\rm rel} = \Omega_{\rm s}/\Omega_{\rm DM}$ to the total DM density, so the limits are applied by
\begin{eqnarray} 
    \frac{1}{d} f_{\rm rel} ~\sigma_{\rm SI} < \sigma_{\rm LZ},
\end{eqnarray}
where $\sigma_{\rm LZ}$ is the 90\% CL limits from the LUX-ZEPLIN experiment~(LZ)~\cite{LZ:2022ufs}, $d$ is the dilution factor. Large $\lambda_{\rm hs}$ can evade DM direct detection due to a  small $f_{\rm rel}$. However, the region where a strong EWPT occurs is fully excluded by the current results of LZ if the dilution factor $d$ is smaller than 2, which is the upper value obtained in ~\cite{Xiao:2022oaq}. Along the green line in Fig.\ref{fig:DM}, we can tune the model parameters to achieve an extremely supercooled EWPT, which results in a larger dilution factor. For example, the region of supercooling with $m_{\rm s}>100\gev$ can evade DM direct detection if $d>5$. 

Therefore, the supercooled EWPT found in this paper can alleviate the tension between the requirement of an EWPT and the constraints from DM direct detection in the SSM. It provides the opportunity to simultaneously explain the GW signals and baryon asymmetry, while satisfying constraints from DM detection and collider searches. 
Meanwhile, the survival parameter space can be cross-checked by future GW and DM detection: a more supercooling case gives a strong GW signal, while a less supercooling case results in a large effective DM-nucleons cross section.

\section{Conclusion}

In this work we investigated the phenomenon of extreme supercooling and the occurrence of a strong first-order electroweak phase transition in the singlet extension of the Standard Model. Our findings revealed that the percolation temperature can significantly and continuously decrease with increasing mixing coupling $\lambda_{\rm hs}$. Consequently, by appropriately tuning the model parameters, we found that it is possible to achieve a percolation temperature of a few GeV. 
We explored the implications of such a phase transition on the generation of a stochastic GW background resulted from bubble collisions at the percolation temperature. The observed signals from pulsar timing array collaborations could be reasonably explained by this GW background, disregarding possible red-shift effects.
If this extreme supercooling and strong first-order electroweak phase transition indeed characterizes the nature of our universe, it will have a profound impact on the dark matter properties. Further research in this direction is warranted to fully understand the implications and consequences of our findings.

~

\section*{Acknowledgments}
This work was supported by the National Natural Science Foundation of China under grant numbers 12105248, 11821505, 12075300 and 12335005, the Peng-Huan-Wu Theoretical Physics Innovation Center under grant number 12047503, the Key R\&D Program of the Ministry of Science and Technology under grant number 2017YFA0402204, and the Key Research Program of the Chinese Academy of Sciences under grant number XDPB15.

\section*{Author contributions}
Jin Min Yang supervised the project. Yang Xiao and Yang Zhang designed the computational framework and performed the calculations. Yang Zhang and Jin Min Yang wrote the manuscript with input from all authors. All authors discussed the results and made contribution equally to the manuscript.

\bibliography{apssamp}

\providecommand{\noopsort}[1]{}\providecommand{\singleletter}[1]{#1}%

\providecommand{\href}[2]{#2}\begingroup\raggedright\begin{thebibliography}{100}

\bibitem{Xu:2023wog}
H.~Xu, S.~Chen, Y.~Guo et~al., \emph{{Searching for the nano-hertz stochastic
  gravitational wave background with the Chinese pulsar timing array data
  release I}}, \href{https://doi.org/10.1088/1674-4527/acdfa5}{\emph{Res Astron
  Astrophys} {\bfseries 23} (2023) 075024}
  [\href{https://arxiv.org/abs/2306.16216}{{\ttfamily 2306.16216}}].

\bibitem{Reardon:2023gzh}
D.~J. Reardon, A.~Zic, R.~M. Shannon et~al., \emph{{Search for an isotropic
  gravitational-wave background with the parkes pulsar timing array}},
  \href{https://doi.org/10.3847/2041-8213/acdd02}{\emph{Astrophys J Lett}
  {\bfseries 951} (2023) } [\href{https://arxiv.org/abs/2306.16215}{{\ttfamily
  2306.16215}}].

\bibitem{Antoniadis:2023ott}
{\scshape EPTA} collaboration, E.~Collaboration, J.~Antoniadis et~al.,
  \emph{{The second data release from the european pulsar timing array III.
  search for gravitational wave signals}},
  \href{https://doi.org/10.1051/0004-6361/202346844}{\emph{Astron Astrophys}
  {\bfseries 678} (2023) A50}
  [\href{https://arxiv.org/abs/2306.16214}{{\ttfamily 2306.16214}}].

\bibitem{NANOGrav:2023gor}
{\scshape NANOGrav} collaboration, N.~Collaboration, G.~Agazie et~al.,
  \emph{{The NANOGrav 15-year data set: evidence for a gravitational-wave
  background}}, \href{https://doi.org/10.3847/2041-8213/acdac6}{\emph{Astrophys
  J Lett} {\bfseries 951} (2023) }
  [\href{https://arxiv.org/abs/2306.16213}{{\ttfamily 2306.16213}}].

\bibitem{Bai:2023cqj}
Y.~Bai, T.-K. Chen and M.~Korwar, \emph{{QCD-Collapsed Domain Walls: QCD Phase
  Transition and Gravitational Wave Spectroscopy}},
  \href{https://arxiv.org/abs/2306.17160}{{\ttfamily 2306.17160}}.

\bibitem{Kitajima:2023cek}
N.~Kitajima, J.~Lee, K.~Murai, F.~Takahashi and W.~Yin, \emph{{Nanohertz
  Gravitational Waves from Axion Domain Walls Coupled to QCD}},
  \href{https://arxiv.org/abs/2306.17146}{{\ttfamily 2306.17146}}.

\bibitem{Yang:2023aak}
J.~Yang, N.~Xie and F.~P. Huang, \emph{{Nano-Hertz stochastic gravitational
  wave background as hints of ultralight axion particles}},
  \href{https://arxiv.org/abs/2306.17113}{{\ttfamily 2306.17113}}.

\bibitem{Megias:2023kiy}
E.~Megias, G.~Nardini and M.~Quiros, \emph{{Pulsar Timing Array Stochastic
  Background from light Kaluza-Klein resonances}},
  \href{https://arxiv.org/abs/2306.17071}{{\ttfamily 2306.17071}}.

\bibitem{Guo:2023hyp}
S.-Y. Guo, M.~Khlopov, X.~Liu et~al., \emph{{Footprints of axion-like particle
  in pulsar timing array data and JWST observations}}, {\emph{arXiv:2306.17022}
  }.

\bibitem{Han:2023olf}
C.~Han, K.-P. Xie, J.~M. Yang and M.~Zhang, \emph{{Self-interacting dark matter
  implied by nano-Hertz gravitational waves}},
  \href{https://arxiv.org/abs/2306.16966}{{\ttfamily 2306.16966}}.

\bibitem{Deng:2023seh}
S.~Deng and L.~Bian, \emph{{Constraining low-scale dark phase transitions with
  cosmological observations}},
  \href{https://arxiv.org/abs/2304.06576}{{\ttfamily 2304.06576}}.

\bibitem{Bian:2022qbh}
L.~Bian, S.~Ge, C.~Li, J.~Shu and J.~Zong, \emph{{Searching for Domain Wall
  Network by Parkes Pulsar Timing Array}},
  \href{https://arxiv.org/abs/2212.07871}{{\ttfamily 2212.07871}}.

\bibitem{Xue:2021gyq}
X.~Xue, L.~Bian, J.~Shu et~al., \emph{{Constraining cosmological phase
  transitions with the parkes pulsar timing array}},
  \href{https://doi.org/10.1103/PhysRevLett.127.251303}{\emph{Phys Rev Lett}
  {\bfseries 127} (2021) 251303}
  [\href{https://arxiv.org/abs/2110.03096}{{\ttfamily 2110.03096}}].

\bibitem{kitajima2023nanohertz}
N.~Kitajima and K.~Nakayama, \emph{Nanohertz gravitational waves from cosmic
  strings and dark photon dark matter},  2023.

\bibitem{lazarides2023superheavy}
G.~Lazarides, R.~Maji and Q.~Shafi, \emph{Superheavy quasi-stable strings and
  walls bounded by strings in the light of nanograv 15 year data},  2023.

\bibitem{yang2023implication}
A.~Yang, J.~Ma, S.~Jiang and F.~P. Huang, \emph{Implication of nano-hertz
  stochastic gravitational wave on dynamical dark matter through a first-order
  phase transition},  2023.

\bibitem{blasi2023axionic}
S.~Blasi, A.~Mariotti, A.~Rase and A.~Sevrin, \emph{Axionic domain walls at
  pulsar timing arrays: Qcd bias and particle friction},  2023.

\bibitem{addazi2023pulsar}
A.~Addazi, Y.-F. Cai, A.~Marciano and L.~Visinelli, \emph{Have pulsar timing
  array methods detected a cosmological phase transition?},  2023.

\bibitem{broadhurst2023binary}
T.~Broadhurst, C.~Chen, T.~Liu and K.-F. Zheng, \emph{Binary supermassive black
  holes orbiting dark matter solitons: From the dual agn in ugc4211 to
  nanohertz gravitational waves},  2023.

\bibitem{Wang:2022akn}
W.~Wang, W.-L. Xu and J.~M. Yang, \emph{{A hidden self-interacting dark matter
  sector with first order cosmological phase transition and gravitational
  wave}},  \href{https://arxiv.org/abs/2209.11408}{{\ttfamily 2209.11408}}.

\bibitem{Wang:2022lxn}
W.~Wang, K.-P. Xie, W.-L. Xu et~al., \emph{{Cosmological phase transitions,
  gravitational waves and self-interacting dark matter in the singlet extension
  of MSSM}}, \href{https://doi.org/10.1140/epjc/s10052-022-11077-3}{\emph{Eur
  Phys J C} {\bfseries 82} (2022) 1120}
  [\href{https://arxiv.org/abs/2204.01928}{{\ttfamily 2204.01928}}].

\bibitem{madge2023primordial}
E.~Madge, E.~Morgante, C.~Puchades-Ib\'a\~nez et~al., \emph{{Primordial
  gravitational waves in the nano-hertz regime and PTA data \textemdash{}
  towards solving the GW inverse problem}},
  \href{https://doi.org/10.1007/JHEP10(2023)171}{\emph{Journal of High Energy
  Physics} {\bfseries 23} (2020) 171}
  [\href{https://arxiv.org/abs/2306.14856}{{\ttfamily 2306.14856}}].

\bibitem{King:2023cgv}
S.~F. King, D.~Marfatia and M.~H. Rahat, \emph{{Towards distinguishing Dirac
  from Majorana neutrino mass with gravitational waves}},
  \href{https://arxiv.org/abs/2306.05389}{{\ttfamily 2306.05389}}.

\bibitem{Chen:2019xse}
Z.-C. Chen, C.~Yuan and Q.-G. Huang, \emph{{Pulsar timing array constraints on
  primordial black holes with NANOGrav 11-year dataset}},
  \href{https://doi.org/10.1103/PhysRevLett.124.251101}{\emph{Phys Rev Lett}
  {\bfseries 124} (2020) 251101}
  [\href{https://arxiv.org/abs/1910.12239}{{\ttfamily 1910.12239}}].

\bibitem{Niu:2023bsr}
X.~Niu and M.~H. Rahat, \emph{{NANOGrav signal from axion inflation}},
  \href{https://arxiv.org/abs/2307.01192}{{\ttfamily 2307.01192}}.

\bibitem{antoniadis2023second}
J.~Antoniadis, P.~Arumugam, S.~Arumugam, P.~Auclair, S.~Babak, M.~Bagchi
  et~al., \emph{The second data release from the european pulsar timing array:
  V. implications for massive black holes, dark matter and the early universe},
   2023.

\bibitem{lu2023nanohertz}
B.-Q. Lu and C.-W. Chiang, \emph{Nano-hertz stochastic gravitational wave
  background from domain wall annihilation},  2023.

\bibitem{Huang:2023chx}
H.-L. Huang, Y.~Cai, J.-Q. Jiang, J.~Zhang and Y.-S. Piao, \emph{{Supermassive
  primordial black holes in multiverse: for nano-Hertz gravitational wave and
  high-redshift JWST galaxies}},
  \href{https://arxiv.org/abs/2306.17577}{{\ttfamily 2306.17577}}.

\bibitem{Jiang:2023gfe}
J.-Q. Jiang, Y.~Cai, G.~Ye and Y.-S. Piao, \emph{{Broken blue-tilted
  inflationary gravitational waves: a joint analysis of NANOGrav 15-year and
  BICEP/Keck 2018 data}},  \href{https://arxiv.org/abs/2307.15547}{{\ttfamily
  2307.15547}}.

\bibitem{Zhu:2023lbf}
M.~Zhu, G.~Ye and Y.~Cai, \emph{{Pulsar timing array observations as possible
  hints for nonsingular cosmology}},
  \href{https://doi.org/10.1140/epjc/s10052-023-11963-4}{\emph{Eur Phys J C}
  {\bfseries 83} (2023) 816}
  [\href{https://arxiv.org/abs/2307.16211}{{\ttfamily 2307.16211}}].

\bibitem{Wang2023ost}
S.~Wang, Z.-C. Zhao, J.-P. Li and Q.-H. Zhu, \emph{{Exploring the Implications
  of 2023 Pulsar Timing Array Datasets for Scalar-Induced Gravitational Waves
  and Primordial Black Holes}},
  \href{https://arxiv.org/abs/2307.00572}{{\ttfamily 2307.00572}}.

\bibitem{Cai:2023dls}
Y.-F. Cai, X.-C. He, X.-H. Ma et~al., \emph{{Limits on scalar-induced
  gravitational waves from the stochastic background by pulsar timing array
  observations}}, \href{https://doi.org/10.1016/j.scib.2023.10.027}{\emph{Sci
  Bull} {\bfseries In Press} (2023) }
  [\href{https://arxiv.org/abs/2306.17822}{{\ttfamily 2306.17822}}].

\bibitem{NANOGrav:2023hvm}
{\scshape NANOGrav} collaboration, N.~Collaboration, A.~Afzal et~al.,
  \emph{{The NANOGrav 15-year data set: search for signals from new physics}},
  \href{https://doi.org/10.3847/2041-8213/acdc91}{\emph{Astrophys J Lett}
  {\bfseries 951} (2023) } [\href{https://arxiv.org/abs/2306.16219}{{\ttfamily
  2306.16219}}].

\bibitem{Pietroni:1992in}
M.~Pietroni, \emph{{The Electroweak phase transition in a nonminimal
  supersymmetric model}},
  \href{https://doi.org/10.1016/0550-3213(93)90635-3}{\emph{Nucl. Phys. B}
  {\bfseries 402} (1993) 27}
  [\href{https://arxiv.org/abs/hep-ph/9207227}{{\ttfamily hep-ph/9207227}}].

\bibitem{Cline:1996mga}
J.~M. Cline and P.-A. Lemieux, \emph{{Electroweak phase transition in two Higgs
  doublet models}}, \href{https://doi.org/10.1103/PhysRevD.55.3873}{\emph{Phys
  Rev D} {\bfseries 55} (1997) 3873}
  [\href{https://arxiv.org/abs/hep-ph/9609240}{{\ttfamily hep-ph/9609240}}].

\bibitem{Ham:2004nv}
S.~W. Ham, S.~K. OH, C.~M. Kim, E.~J. Yoo and D.~Son, \emph{{Electroweak phase
  transition in a nonminimal supersymmetric model}},
  \href{https://doi.org/10.1103/PhysRevD.70.075001}{\emph{Phys Rev D}
  {\bfseries 70} (2004) 075001}
  [\href{https://arxiv.org/abs/hep-ph/0406062}{{\ttfamily hep-ph/0406062}}].

\bibitem{Funakubo:2005pu}
K.~Funakubo, S.~Tao and F.~Toyoda, \emph{{Phase transitions in the NMSSM}},
  \href{https://doi.org/10.1143/PTP.114.369}{\emph{Prog Theor Phys} {\bfseries
  114} (2005) 369} [\href{https://arxiv.org/abs/hep-ph/0501052}{{\ttfamily
  hep-ph/0501052}}].

\bibitem{Barger:2008jx}
V.~Barger, P.~Langacker, M.~McCaskey, M.~Ramsey-Musolf and G.~Shaughnessy,
  \emph{{Complex Singlet Extension of the Standard Model}},
  \href{https://doi.org/10.1103/PhysRevD.79.015018}{\emph{Phys Rev D}
  {\bfseries 79} (2009) 015018}
  [\href{https://arxiv.org/abs/0811.0393}{{\ttfamily 0811.0393}}].

\bibitem{Chung:2010cd}
D.~J.~H. Chung and A.~J. Long, \emph{{Electroweak Phase Transition in the
  $\mu\nu$SSM}}, \href{https://doi.org/10.1103/PhysRevD.81.123531}{\emph{Phys
  Rev D} {\bfseries 81} (2010) 123531}
  [\href{https://arxiv.org/abs/1004.0942}{{\ttfamily 1004.0942}}].

\bibitem{Espinosa:2011ax}
J.~R. Espinosa, T.~Konstandin and F.~Riva, \emph{{Strong Electroweak Phase
  Transitions in the Standard Model with a Singlet}},
  \href{https://doi.org/10.1016/j.nuclphysb.2011.09.010}{\emph{Nucl. Phys. B}
  {\bfseries 854} (2012) 592}
  [\href{https://arxiv.org/abs/1107.5441}{{\ttfamily 1107.5441}}].

\bibitem{Chowdhury:2011ga}
T.~A. Chowdhury, M.~Nemevsek, G.~Senjanovic and Y.~Zhang, \emph{{Dark matter as
  the trigger of strong electroweak phase transition}},
  \href{https://doi.org/10.1088/1475-7516/2012/02/029}{\emph{Journal of
  Cosmology and Astroparticle Physics} {\bfseries 02} (2012) 029}
  [\href{https://arxiv.org/abs/1110.5334}{{\ttfamily 1110.5334}}].

\bibitem{Gil:2012ya}
G.~Gil, P.~Chankowski and M.~Krawczyk, \emph{{Inert dark matter and strong
  electroweak phase transition}},
  \href{https://doi.org/10.1016/j.physletb.2012.09.052}{\emph{Phys. Lett. B}
  {\bfseries 717} (2012) 396}
  [\href{https://arxiv.org/abs/1207.0084}{{\ttfamily 1207.0084}}].

\bibitem{Carena:2012np}
M.~Carena, G.~Nardini, M.~Quiros and C.~E. Wagner, \emph{{MSSM Electroweak
  Baryogenesis and LHC Data}}, \href{https://doi.org/10.1007/Journal of High
  Energy Physics02(2013)001}{\emph{Journal of High Energy Physics} {\bfseries
  02} (2013) 001} [\href{https://arxiv.org/abs/1207.6330}{{\ttfamily
  1207.6330}}].

\bibitem{No:2013wsa}
J.~M. No and M.~Ramsey-Musolf, \emph{{Probing the Higgs Portal at the LHC
  Through Resonant di-Higgs Production}},
  \href{https://doi.org/10.1103/PhysRevD.89.095031}{\emph{Phys Rev D}
  {\bfseries 89} (2014) 095031}
  [\href{https://arxiv.org/abs/1310.6035}{{\ttfamily 1310.6035}}].

\bibitem{Dorsch:2013wja}
G.~C. Dorsch, S.~J. Huber and J.~M. No, \emph{{A strong electroweak phase
  transition in the 2HDM after LHC8}}, \href{https://doi.org/10.1007/Journal of
  High Energy Physics10(2013)029}{\emph{Journal of High Energy Physics}
  {\bfseries 10} (2013) 029} [\href{https://arxiv.org/abs/1305.6610}{{\ttfamily
  1305.6610}}].

\bibitem{Curtin:2014jma}
D.~Curtin, P.~Meade and C.-T. Yu, \emph{{Testing Electroweak Baryogenesis with
  Future Colliders}}, \href{https://doi.org/10.1007/Journal of High Energy
  Physics11(2014)127}{\emph{Journal of High Energy Physics} {\bfseries 11}
  (2014) 127} [\href{https://arxiv.org/abs/1409.0005}{{\ttfamily 1409.0005}}].

\bibitem{Huang:2014ifa}
W.~Huang, Z.~Kang, J.~Shu et~al., \emph{{New insights in the electroweak phase
  transition in the NMSSM}},
  \href{https://doi.org/10.1103/PhysRevD.91.025006}{\emph{Phys Rev D}
  {\bfseries 91} (2015) 025006}
  [\href{https://arxiv.org/abs/1405.1152}{{\ttfamily 1405.1152}}].

\bibitem{Profumo:2014opa}
S.~Profumo, M.~J. Ramsey-Musolf, C.~L. Wainwright et~al.,
  \emph{{Singlet-catalyzed electroweak phase transitions and precision higgs
  boson studies}}, \href{https://doi.org/10.1103/PhysRevD.91.035018}{\emph{Phys
  Rev D} {\bfseries 91} (2015) 035018}
  [\href{https://arxiv.org/abs/1407.5342}{{\ttfamily 1407.5342}}].

\bibitem{Kozaczuk:2014kva}
J.~Kozaczuk, S.~Profumo, L.~S. Haskins and C.~L. Wainwright,
  \emph{{Cosmological Phase Transitions and their Properties in the NMSSM}},
  \href{https://doi.org/10.1007/Journal of High Energy
  Physics01(2015)144}{\emph{Journal of High Energy Physics} {\bfseries 01}
  (2015) 144} [\href{https://arxiv.org/abs/1407.4134}{{\ttfamily 1407.4134}}].

\bibitem{Jiang:2015cwa}
M.~Jiang, L.~Bian, W.~Huang and J.~Shu, \emph{{Impact of a complex singlet:
  Electroweak baryogenesis and dark matter}},
  \href{https://doi.org/10.1103/PhysRevD.93.065032}{\emph{Phys Rev D}
  {\bfseries 93} (2016) 065032}
  [\href{https://arxiv.org/abs/1502.07574}{{\ttfamily 1502.07574}}].

\bibitem{Curtin:2016urg}
D.~Curtin, P.~Meade and H.~Ramani, \emph{{Thermal Resummation and Phase
  Transitions}},
  \href{https://doi.org/10.1140/epjc/s10052-018-6268-0}{\emph{Eur Phys J C}
  {\bfseries 78} (2018) 787}
  [\href{https://arxiv.org/abs/1612.00466}{{\ttfamily 1612.00466}}].

\bibitem{Vaskonen:2016yiu}
V.~Vaskonen, \emph{{Electroweak baryogenesis and gravitational waves from a
  real scalar singlet}},
  \href{https://doi.org/10.1103/PhysRevD.95.123515}{\emph{Phys Rev D}
  {\bfseries 95} (2017) 123515}
  [\href{https://arxiv.org/abs/1611.02073}{{\ttfamily 1611.02073}}].

\bibitem{Dorsch:2016nrg}
G.~Dorsch, S.~Huber, T.~Konstandin et~al., \emph{{A second higgs doublet in the
  early universe: baryogenesis and gravitational waves}},
  \href{https://doi.org/10.1088/1475-7516/2017/05/052}{\emph{Journal of
  Cosmology and Astroparticle Physics} {\bfseries 05} (2017) 052}
  [\href{https://arxiv.org/abs/1611.05874}{{\ttfamily 1611.05874}}].

\bibitem{Huang:2016cjm}
P.~Huang, A.~J. Long and L.-T. Wang, \emph{{Probing the electroweak phase
  transition with higgs factories and gravitational waves}},
  \href{https://doi.org/10.1103/PhysRevD.94.075008}{\emph{Phys Rev D}
  {\bfseries 94} (2016) 075008}
  [\href{https://arxiv.org/abs/1608.06619}{{\ttfamily 1608.06619}}].

\bibitem{Chala:2016ykx}
M.~Chala, G.~Nardini and I.~Sobolev, \emph{{Unified explanation for dark matter
  and electroweak baryogenesis with direct detection and gravitational wave
  signatures}}, \href{https://doi.org/10.1103/PhysRevD.94.055006}{\emph{Phys
  Rev D} {\bfseries 94} (2016) 055006}
  [\href{https://arxiv.org/abs/1605.08663}{{\ttfamily 1605.08663}}].

\bibitem{Basler:2016obg}
P.~Basler, M.~Krause, M.~Muhlleitner, J.~Wittbrodt and A.~Wlotzka,
  \emph{{Strong First Order Electroweak Phase Transition in the CP-Conserving
  2HDM Revisited}}, \href{https://doi.org/10.1007/Journal of High Energy
  Physics02(2017)121}{\emph{Journal of High Energy Physics} {\bfseries 02}
  (2017) 121} [\href{https://arxiv.org/abs/1612.04086}{{\ttfamily
  1612.04086}}].

\bibitem{Beniwal:2017eik}
A.~Beniwal, M.~Lewicki, J.~D. Wells, M.~White and A.~G. Williams,
  \emph{{Gravitational wave, collider and dark matter signals from a scalar
  singlet electroweak baryogenesis}}, \href{https://doi.org/10.1007/Journal of
  High Energy Physics08(2017)108}{\emph{Journal of High Energy Physics}
  {\bfseries 08} (2017) 108}
  [\href{https://arxiv.org/abs/1702.06124}{{\ttfamily 1702.06124}}].

\bibitem{Bernon:2017jgv}
J.~Bernon, L.~Bian and Y.~Jiang, \emph{{A new insight into the phase transition
  in the early Universe with two Higgs doublets}},
  \href{https://doi.org/10.1007/Journal of High Energy
  Physics05(2018)151}{\emph{Journal of High Energy Physics} {\bfseries 05}
  (2018) 151} [\href{https://arxiv.org/abs/1712.08430}{{\ttfamily
  1712.08430}}].

\bibitem{Kurup:2017dzf}
G.~Kurup and M.~Perelstein, \emph{{Dynamics of Electroweak Phase Transition In
  Singlet-Scalar Extension of the Standard Model}},
  \href{https://doi.org/10.1103/PhysRevD.96.015036}{\emph{Phys Rev D}
  {\bfseries 96} (2017) 015036}
  [\href{https://arxiv.org/abs/1704.03381}{{\ttfamily 1704.03381}}].

\bibitem{Andersen:2017ika}
J.~O. Andersen, T.~Gorda, A.~Helset, L.~Niemi, T.~V.~I. Tenkanen, A.~Tranberg
  et~al., \emph{{Nonperturbative Analysis of the Electroweak Phase Transition
  in the Two Higgs Doublet Model}},
  \href{https://doi.org/10.1103/PhysRevLett.121.191802}{\emph{Phys. Rev. Lett.}
  {\bfseries 121} (2018) 191802}
  [\href{https://arxiv.org/abs/1711.09849}{{\ttfamily 1711.09849}}].

\bibitem{Chiang:2017nmu}
C.-W. Chiang, M.~J. Ramsey-Musolf and E.~Senaha, \emph{{Standard model with a
  complex scalar singlet: cosmological implications and theoretical
  considerations}},
  \href{https://doi.org/10.1103/PhysRevD.97.015005}{\emph{Phys Rev D}
  {\bfseries 97} (2018) 015005}
  [\href{https://arxiv.org/abs/1707.09960}{{\ttfamily 1707.09960}}].

\bibitem{Dorsch:2017nza}
G.~C. Dorsch, S.~J. Huber, K.~Mimasu and J.~M. No, \emph{{The Higgs Vacuum
  Uplifted: Revisiting the Electroweak Phase Transition with a Second Higgs
  Doublet}}, \href{https://doi.org/10.1007/Journal of High Energy
  Physics12(2017)086}{\emph{Journal of High Energy Physics} {\bfseries 12}
  (2017) 086} [\href{https://arxiv.org/abs/1705.09186}{{\ttfamily
  1705.09186}}].

\bibitem{Beniwal:2018hyi}
A.~Beniwal, M.~Lewicki, M.~White and A.~G. Williams, \emph{{Gravitational waves
  and electroweak baryogenesis in a global study of the extended scalar singlet
  model}}, \href{https://doi.org/10.1007/Journal of High Energy
  Physics02(2019)183}{\emph{Journal of High Energy Physics} {\bfseries 02}
  (2019) 183} [\href{https://arxiv.org/abs/1810.02380}{{\ttfamily
  1810.02380}}].

\bibitem{Alves:2018jsw}
A.~Alves, T.~Ghosh, H.-K. Guo, K.~Sinha and D.~Vagie, \emph{{Collider and
  Gravitational Wave Complementarity in Exploring the Singlet Extension of the
  Standard Model}}, \href{https://doi.org/10.1007/Journal of High Energy
  Physics04(2019)052}{\emph{Journal of High Energy Physics} {\bfseries 04}
  (2019) 052} [\href{https://arxiv.org/abs/1812.09333}{{\ttfamily
  1812.09333}}].

\bibitem{Bruggisser:2018mrt}
S.~Bruggisser, B.~Von~Harling, O.~Matsedonskyi and G.~Servant,
  \emph{{Electroweak Phase Transition and Baryogenesis in Composite Higgs
  Models}}, \href{https://doi.org/10.1007/Journal of High Energy
  Physics12(2018)099}{\emph{Journal of High Energy Physics} {\bfseries 12}
  (2018) 099} [\href{https://arxiv.org/abs/1804.07314}{{\ttfamily
  1804.07314}}].

\bibitem{Athron:2019teq}
P.~Athron, C.~Balazs, A.~Fowlie et~al., \emph{{Strong first-order phase
  transitions in the NMSSM \textemdash{} a comprehensive survey}},
  \href{https://doi.org/10.1007/Journal of High Energy
  Physics11(2019)151}{\emph{Journal of High Energy Physics} {\bfseries 11}
  (2019) 151} [\href{https://arxiv.org/abs/1908.11847}{{\ttfamily
  1908.11847}}].

\bibitem{Kainulainen:2019kyp}
K.~Kainulainen, V.~Keus, L.~Niemi, K.~Rummukainen, T.~V.~I. Tenkanen and
  V.~Vaskonen, \emph{{On the validity of perturbative studies of the
  electroweak phase transition in the Two Higgs Doublet model}},
  \href{https://doi.org/10.1007/Journal of High Energy
  Physics06(2019)075}{\emph{Journal of High Energy Physics} {\bfseries 06}
  (2019) 075} [\href{https://arxiv.org/abs/1904.01329}{{\ttfamily
  1904.01329}}].

\bibitem{Bian:2019kmg}
L.~Bian, Y.~Wu and K.-P. Xie, \emph{{Electroweak phase transition with
  composite Higgs models: calculability, gravitational waves and collider
  searches}}, \href{https://doi.org/10.1007/Journal of High Energy
  Physics12(2019)028}{\emph{Journal of High Energy Physics} {\bfseries 12}
  (2019) 028} [\href{https://arxiv.org/abs/1909.02014}{{\ttfamily
  1909.02014}}].

\bibitem{Li:2019tfd}
H.-L. Li, M.~Ramsey-Musolf and S.~Willocq, \emph{{Probing a scalar
  singlet-catalyzed electroweak phase transition with resonant di-Higgs boson
  production in the $4b$ channel}},
  \href{https://doi.org/10.1103/PhysRevD.100.075035}{\emph{Phys Rev D}
  {\bfseries 100} (2019) 075035}
  [\href{https://arxiv.org/abs/1906.05289}{{\ttfamily 1906.05289}}].

\bibitem{Chiang:2019oms}
C.-W. Chiang and B.-Q. Lu, \emph{{First-order electroweak phase transition in a
  complex singlet model with $\mathbb{Z}_3$ symmetry}},
  \href{https://doi.org/10.1007/Journal of High Energy
  Physics07(2020)082}{\emph{Journal of High Energy Physics} {\bfseries 07}
  (2020) 082} [\href{https://arxiv.org/abs/1912.12634}{{\ttfamily
  1912.12634}}].

\bibitem{Xie:2020bkl}
K.-P. Xie, L.~Bian and Y.~Wu, \emph{{Electroweak baryogenesis and gravitational
  waves in a composite higgs model with high dimensional fermion
  representations}}, \href{https://doi.org/10.1007/Journal of High Energy
  Physics12(2020)047}{\emph{Journal of High Energy Physics} {\bfseries 12}
  (2020) 047} [\href{https://arxiv.org/abs/2005.13552}{{\ttfamily
  2005.13552}}].

\bibitem{Azatov:2022tii}
A.~Azatov, G.~Barni, S.~Chakraborty et~al., \emph{{Ultra-relativistic bubbles
  from the simplest higgs portal and their cosmological consequences}},
  \href{https://doi.org/10.1007/Journal of High Energy
  Physics10(2022)017}{\emph{Journal of High Energy Physics} {\bfseries 10}
  (2022) 017} [\href{https://arxiv.org/abs/2207.02230}{{\ttfamily
  2207.02230}}].

\bibitem{Bell:2020gug}
N.~F. Bell, M.~J. Dolan, L.~S. Friedrich, M.~J. Ramsey-Musolf and R.~R. Volkas,
  \emph{{Two-Step Electroweak Symmetry-Breaking: Theory Meets Experiment}},
  \href{https://doi.org/10.1007/Journal of High Energy
  Physics05(2020)050}{\emph{Journal of High Energy Physics} {\bfseries 05}
  (2020) 050} [\href{https://arxiv.org/abs/2001.05335}{{\ttfamily
  2001.05335}}].

\bibitem{Han:2020ekm}
X.-F. Han, L.~Wang and Y.~Zhang, \emph{{Dark matter, electroweak phase
  transition, and gravitational waves in the type II two-higgs-doublet model
  with a singlet scalar field}},
  \href{https://doi.org/10.1103/PhysRevD.103.035012}{\emph{Phys Rev D}
  {\bfseries 103} (2021) 035012}
  [\href{https://arxiv.org/abs/2010.03730}{{\ttfamily 2010.03730}}].

\bibitem{Ghosh:2022fzp}
P.~Ghosh, T.~Ghosh and S.~Roy, \emph{{Interplay among gravitational waves, dark
  matter and collider signals in the singlet scalar extended type-II seesaw
  model}},  \href{https://arxiv.org/abs/2211.15640}{{\ttfamily 2211.15640}}.

\bibitem{Cao:2022ocg}
Q.-H. Cao, K.~Hashino, X.-X. Li et~al., \emph{{Multi-step phase transition and
  gravitational wave from general $\mathbb{Z}_2$ scalar extensions}},
  {\emph{arXiv:2212.07756} }.

\bibitem{Zhao:2022cnn}
Z.~Zhao, Y.~Di, L.~Bian et~al., \emph{{Probing the electroweak symmetry
  breaking history with gravitational waves}}, {\emph{arXiv:2204.04427} }.

\bibitem{Chatterjee:2022pxf}
A.~Chatterjee, A.~Datta and S.~Roy, \emph{{Electroweak phase transition in the
  Z$_{3}$-invariant NMSSM: Implications of LHC and Dark matter searches and
  prospects of detecting the gravitational waves}},
  \href{https://doi.org/10.1007/Journal of High Energy
  Physics06(2022)108}{\emph{Journal of High Energy Physics} {\bfseries 06}
  (2022) 108} [\href{https://arxiv.org/abs/2202.12476}{{\ttfamily
  2202.12476}}].

\bibitem{ashoorioon2009strong}
A.~Ashoorioon and T.~Konstandin, \emph{Strong electroweak phase transitions
  without collider traces}, {\emph{Journal of High Energy Physics} {\bfseries
  2009} (2009) 086}.

\bibitem{Baratella:2018pxi}
P.~Baratella, A.~Pomarol and F.~Rompineve, \emph{{The supercooled universe}},
  \href{https://doi.org/10.1007/Journal of High Energy
  Physics03(2019)100}{\emph{Journal of High Energy Physics} {\bfseries 03}
  (2019) 100} [\href{https://arxiv.org/abs/1812.06996}{{\ttfamily
  1812.06996}}].

\bibitem{Lewicki:2020jiv}
M.~Lewicki and V.~Vaskonen, \emph{{Gravitational wave spectra from strongly
  supercooled phase transitions}},
  \href{https://doi.org/10.1140/epjc/s10052-020-08589-1}{\emph{Eur Phys J C}
  {\bfseries 80} (2020) 1003}
  [\href{https://arxiv.org/abs/2007.04967}{{\ttfamily 2007.04967}}].

\bibitem{Wang:2020jrd}
X.~Wang, F.~P. Huang and X.~Zhang, \emph{{Phase transition dynamics and
  gravitational wave spectra of strong first-order phase transition in
  supercooled universe}},
  \href{https://doi.org/10.1088/1475-7516/2020/05/045}{\emph{Journal of
  Cosmology and Astroparticle Physics} {\bfseries 05} (2020) 045}
  [\href{https://arxiv.org/abs/2003.08892}{{\ttfamily 2003.08892}}].

\bibitem{Athron:2022mmm}
P.~Athron, C.~Bal\'azs and L.~Morris, \emph{{Supercool subtleties of
  cosmological phase transitions}},
  \href{https://doi.org/10.1088/1475-7516/2023/03/006}{\emph{Journal of
  Cosmology and Astroparticle Physics} {\bfseries 03} (2023) 006}
  [\href{https://arxiv.org/abs/2212.07559}{{\ttfamily 2212.07559}}].

\bibitem{Kobakhidze:2017mru}
A.~Kobakhidze, C.~Lagger, A.~Manning et~al., \emph{{Gravitational waves from a
  supercooled electroweak phase transition and their detection with pulsar
  timing arrays}},
  \href{https://doi.org/10.1140/epjc/s10052-017-5132-y}{\emph{Eur Phys J C}
  {\bfseries 77} (2017) 570}
  [\href{https://arxiv.org/abs/1703.06552}{{\ttfamily 1703.06552}}].

\bibitem{Cai:2017tmh}
R.-G. Cai, M.~Sasaki and S.-J. Wang, \emph{{The gravitational waves from the
  first-order phase transition with a dimension-six operator}},
  \href{https://doi.org/10.1088/1475-7516/2017/08/004}{\emph{Journal of
  Cosmology and Astroparticle Physics} {\bfseries 08} (2017) 004}
  [\href{https://arxiv.org/abs/1707.03001}{{\ttfamily 1707.03001}}].

\bibitem{Cline:2013gha}
J.~M. Cline, K.~Kainulainen, P.~Scott et~al., \emph{{Update on scalar singlet
  dark matter}}, \href{https://doi.org/10.1103/PhysRevD.88.055025}{\emph{Phys
  Rev D} {\bfseries 88} (2013) 055025}
  [\href{https://arxiv.org/abs/1306.4710}{{\ttfamily 1306.4710}}].

\bibitem{GAMBIT:2017gge}
{\scshape GAMBIT} collaboration, G.~Collaboration, P.~Athron et~al.,
  \emph{{Status of the scalar singlet dark matter model}},
  \href{https://doi.org/10.1140/epjc/s10052-017-5113-1}{\emph{Eur Phys J C}
  {\bfseries 77} (2017) 568}
  [\href{https://arxiv.org/abs/1705.07931}{{\ttfamily 1705.07931}}].

\bibitem{Xiao:2022oaq}
Y.~Xiao, J.~M. Yang and Y.~Zhang, \emph{{Dilution of dark matter relic density
  in singlet extension models}}, \href{https://doi.org/10.1007/Journal of High
  Energy Physics02(2023)008}{\emph{Journal of High Energy Physics} {\bfseries
  02} (2023) 008} [\href{https://arxiv.org/abs/2207.14519}{{\ttfamily
  2207.14519}}].

\bibitem{Roy:2022gop}
S.~Roy, \emph{{Dilution of dark matter Relic abundance due to First Order
  Electroweak Phase Transition in the singlet scalar extended type-II seesaw
  model}},  \href{https://arxiv.org/abs/2212.11230}{{\ttfamily 2212.11230}}.

\bibitem{Hambye:2018qjv}
T.~Hambye, A.~Strumia and D.~Teresi, \emph{{Super-cool dark matter}},
  \href{https://doi.org/10.1007/Journal of High Energy
  Physics08(2018)188}{\emph{Journal of High Energy Physics} {\bfseries 08}
  (2018) 188} [\href{https://arxiv.org/abs/1805.01473}{{\ttfamily
  1805.01473}}].

\bibitem{Baldes:2021aph}
I.~Baldes, Y.~Gouttenoire, F.~Sala et~al., \emph{{Supercool composite dark
  matter beyond 100 TeV}}, \href{https://doi.org/10.1007/Journal of High Energy
  Physics07(2022)084}{\emph{Journal of High Energy Physics} {\bfseries 07}
  (2022) 084} [\href{https://arxiv.org/abs/2110.13926}{{\ttfamily
  2110.13926}}].

\bibitem{Caprini:2019egz}
C.~Caprini, M.~Chala, G.~C. Dorsch et~al., \emph{{Detecting gravitational waves
  from cosmological phase transitions with LISA: an update}},
  \href{https://doi.org/10.1088/1475-7516/2020/03/024}{\emph{Journal of
  Cosmology and Astroparticle Physics} {\bfseries 03} (2020) 024}
  [\href{https://arxiv.org/abs/1910.13125}{{\ttfamily 1910.13125}}].

\bibitem{Athron:2023xlk}
P.~Athron, C.~Bal\'azs, A.~Fowlie et~al., \emph{{Cosmological phase
  transitions: from perturbative particle physics to gravitational waves}},
  {\emph{arXiv:2305.02357} }.

\bibitem{parwani1992resummation}
R.~R. Parwani, \emph{Resummation in a hot scalar field theory}, {\emph{Phys Rev
  D} {\bfseries 45} (1992) 4695}.

\bibitem{Croon_2021}
D.~Croon, O.~Gould, P.~Schicho, T.~V.~I. Tenkanen and G.~White,
  \emph{Theoretical uncertainties for cosmological first-order phase
  transitions}, \href{https://doi.org/10.1007/Journal of High Energy
  Physics04(2021)055}{\emph{Journal of High Energy Physics} {\bfseries 2021}
  (2021) }.

\bibitem{Niemi_2021}
L.~Niemi, P.~Schicho and T.~V. Tenkanen, \emph{Singlet-assisted electroweak
  phase transition at two loops},
  \href{https://doi.org/10.1103/physrevd.103.115035}{\emph{Phys Rev D}
  {\bfseries 103} (2021) }.

\bibitem{Schicho_2021}
P.~M. Schicho, T.~V.~I. Tenkanen and J.~Österman, \emph{Robust approach to
  thermal resummation: standard model meets a singlet},
  \href{https://doi.org/10.1007/Journal of High Energy
  Physics06(2021)130}{\emph{Journal of High Energy Physics} {\bfseries 2021}
  (2021) }.

\bibitem{Ekstedt_2023}
A.~Ekstedt, P.~Schicho and T.~V. Tenkanen, \emph{{DRalgo}: a package for
  effective field theory approach for thermal phase transitions},
  \href{https://doi.org/10.1016/j.cpc.2023.108725}{\emph{Comput Phys Commun}
  {\bfseries 288} (2023) 108725}.

\bibitem{athron2022arbitrary}
P.~Athron, C.~Balazs, A.~Fowlie et~al., \emph{{How arbitrary are perturbative
  calculations of the electroweak phase transition?}},
  \href{https://doi.org/10.1007/JHEP01(2023)050}{\emph{Journal of High Energy
  Physics} {\bfseries 01} (2023) 050}
  [\href{https://arxiv.org/abs/2208.01319}{{\ttfamily 2208.01319}}].

\bibitem{linde1983decay}
A.~D. Linde, \emph{Decay of the false vacuum at finite temperature},
  {\emph{Nucl Phys B} {\bfseries 216} (1983) 421}.

\bibitem{rubakov2009classical}
V.~Rubakov, \emph{Classical theory of gauge fields}. Princeton University
  Press, 2009.

\bibitem{guth1981cosmological}
A.~H. Guth and E.~J. Weinberg, \emph{Cosmological consequences of a first-order
  phase transition in the su(5) grand unified model}, {\emph{Phys Rev D}
  {\bfseries 23} (1981) 876}.

\bibitem{hindmarsh2019gravitational}
M.~Hindmarsh and M.~Hijazi, \emph{Gravitational waves from first order
  cosmological phase transitions in the sound shell model}, {\emph{Journal of
  Cosmology and Astroparticle Physics} {\bfseries 2019} (2019) 062}.

\bibitem{quiros1998finite}
M.~Quiros, \emph{Finite temperature field theory and phase transitions},
  {\emph{Proceedings, Summer school in high-energy physics and cosmology:
  Trieste, Italy} {\bfseries 1999} (1998) 187}.

\bibitem{wainwright2012cosmotransitions}
C.~L. Wainwright, \emph{Cosmotransitions: computing cosmological phase
  transition temperatures and bubble profiles with multiple fields},
  {\emph{Comput Phys Commun} {\bfseries 183} (2012) 2006}.

\bibitem{Balazs:2023kuk}
C.~Bal\'azs, Y.~Xiao, J.~M. Yang et~al., \emph{{New vacuum stability limit from
  cosmological history}}, {\emph{arXiv:2301.09283} }.

\bibitem{Shang:2023gfy}
L.~Shang and Y.~Zhang, \emph{{EasyScan\_HEP: a tool for connecting programs to
  scan the parameter space of physics models}}, {\emph{arXiv:2304.03636} }.

\bibitem{hindmarsh2014gravitational}
M.~Hindmarsh, S.~J. Huber, K.~Rummukainen et~al., \emph{Gravitational waves
  from the sound of a first order phase transition}, {\emph{Phys Rev Lett}
  {\bfseries 112} (2014) 041301}.

\bibitem{huber2008gravitational}
S.~J. Huber and T.~Konstandin, \emph{Gravitational wave production by
  collisions: more bubbles}, {\emph{Journal of Cosmology and Astroparticle
  Physics} {\bfseries 2008} (2008) 022}.

\bibitem{Freese:2022qrl}
K.~Freese and M.~W. Winkler, \emph{{Have pulsar timing arrays detected the hot
  big bang: gravitational waves from strong first order phase transitions in
  the early universe}},
  \href{https://doi.org/10.1103/PhysRevD.106.103523}{\emph{Phys Rev D}
  {\bfseries 106} (2022) 103523}
  [\href{https://arxiv.org/abs/2208.03330}{{\ttfamily 2208.03330}}].

\bibitem{amaro2017laser}
L.~Collaboration, P.~Amaro-Seoane et~al., \emph{Laser interferometer space
  antenna}, {\emph{arXiv:1702.00786} }.

\bibitem{ruan2020taiji}
W.-H. Ruan, Z.-K. Guo, R.-G. Cai et~al., \emph{Taiji program:
  gravitational-wave sources}, {\emph{Int J Mod Phys A} {\bfseries 35} (2020)
  2050075}.

\bibitem{luo2016tianqin}
J.~Luo, L.-S. Chen, H.-Z. Duan et~al., \emph{Tianqin: a space-borne
  gravitational wave detector}, {\emph{Classical and Quantum Gravity}
  {\bfseries 33} (2016) 035010}.

\bibitem{Bian:2021ini}
L.~Bian, R.-G. Cai, S.~Cao et~al., \emph{{The gravitational-wave physics II:
  progress}}, \href{https://doi.org/10.1007/s11433-021-1781-x}{\emph{Sci China
  Phys Mech Astron} {\bfseries 64} (2021) 120401}
  [\href{https://arxiv.org/abs/2106.10235}{{\ttfamily 2106.10235}}].

\bibitem{Zhao:2020iew}
Y.~Zhao and Y.~Lu, \emph{{Stochastic gravitational wave background and
  eccentric stellar compact binaries}},
  \href{https://doi.org/10.1093/mnras/staa2707}{\emph{Mon Not Roy Astron Soc}
  {\bfseries 500} (2020) 1421}
  [\href{https://arxiv.org/abs/2009.01436}{{\ttfamily 2009.01436}}].

\bibitem{Athron:2023mer}
P.~Athron, A.~Fowlie, C.-T. Lu et~al., \emph{{Can supercooled phase transitions
  explain the gravitational wave background observed by pulsar timing
  arrays?}}, {\emph{arXiv:2306.17239} }.

\bibitem{Ertas:2021xeh}
F.~Ertas, F.~Kahlhoefer and C.~Tasillo, \emph{{Turn up the volume: listening to
  phase transitions in hot dark sectors}},
  \href{https://doi.org/10.1088/1475-7516/2022/02/014}{\emph{Journal of
  Cosmology and Astroparticle Physics} {\bfseries 02} (2022) 014}
  [\href{https://arxiv.org/abs/2109.06208}{{\ttfamily 2109.06208}}].

\bibitem{wainwright2009impact}
C.~Wainwright and S.~Profumo, \emph{Impact of a strongly first-order phase
  transition on the abundance of thermal relics}, {\emph{Phys Rev D} {\bfseries
  80} (2009) 103517}.

\bibitem{Ellis:2018mja}
J.~Ellis, M.~Lewicki and J.~M. No, \emph{{On the maximal strength of a
  first-order electroweak phase transition and its gravitational wave signal}},
  \href{https://doi.org/10.1088/1475-7516/2019/04/003}{\emph{Journal of
  Cosmology and Astroparticle Physics} {\bfseries 04} (2019) 003}
  [\href{https://arxiv.org/abs/1809.08242}{{\ttfamily 1809.08242}}].

\bibitem{megevand2008supercooling}
A.~Megevand and A.~D. Sanchez, \emph{Supercooling and phase coexistence in
  cosmological phase transitions}, {\emph{Phys Rev D} {\bfseries 77} (2008)
  063519}.

\bibitem{LZ:2022ufs}
{\scshape LZ} collaboration, L.~Collaboration, J.~Aalbers et~al., \emph{{First
  dark matter search results from the LUX-ZEPLIN (LZ) experiment}},
  \href{https://doi.org/10.1103/PhysRevLett.131.041002}{\emph{Phys Rev Lett}
  {\bfseries 131} (2023) 041002}
  [\href{https://arxiv.org/abs/2207.03764}{{\ttfamily 2207.03764}}].

\end{thebibliography}\endgroup
\bibliographystyle{CitationStyle}

\end{document}